\documentclass[%
reprint,
superscriptaddress,
 amsmath,amssymb,
 aps,
 pra,
eqsecnum
]{revtex4-1}

\usepackage{graphicx}
\usepackage{dcolumn}
\usepackage{bm}
\usepackage{braket}
\usepackage{color}

\def \tr {{\rm Tr}}

\def \hjc {H_{\rm JC}}

\begin{document}


\title{Analytic approach to dynamics of the resonant and off-resonant Jaynes-Cummings systems with cavity losses}

\author{Soshun Ozaki}
\email{ozaki@hosi.phys.s.u-tokyo.ac.jp}
\affiliation{
  Department of Physics, University of Tokyo, Bunkyo, Tokyo 113-0033, Japan}
\author{Hiromichi Nakazato}%
 \affiliation{Department of Physics, Waseda University, Shinjuku, Tokyo 169-8555, Japan}




\date{\today}

\begin{abstract}
A new analytic approach to investigate the zero-temperature time evolution of the Jaynes-Cummings system 
with cavity losses is developed.
With the realistic coupling between the cavity and the environment assumed, 
a simple master equation is derived, leading to 
the explicit analytic solution for the resonant case.
This solution is suitable for the analyses not only on the single excitation states 
but also on many excitation states, which enables us to investigate 
the photon coherent state and to observe sharp collapses and revivals 
under dissipation.
%
For the off-resonant case, on the other hand, 
the present study presents an analytic, systematic method instead.
We examine the small and large detuning limits and 
discuss the condition where the widely-used phenomenological treatment 
is justified.
Explicit evaluations of the time evolutions for various initial states
with finite detuning are also presented.
\end{abstract}

\maketitle


\section{Introduction}
The Jaynes-Cummings (JC) model \cite{jc} is one of the simplest models 
for matters interacting with a quantized mode of the 
electromagnetic field
and contains fertile physics such as the spontaneous emission, 
the Rabi oscillation, and collapses and revivals of the atomic-state probabilities
\cite{jc,shoreknight, puri, eberly}.
These phenomena have been observed in the experiments in optical cavities
\cite{raimond, walther, wineland, leibfried}.
The JC model is now becoming applied to quantum informatics as a way of realization 
of the controlled NOT gate, which plays indispensable role in this field
\cite{nielsen,yanghong, ionicioiu, mishuck, azuma}.

From the viewpoint of experiments in cavities, the noises due to the interaction between the system 
that we focus on and the environment, such as photon losses, are inevitable,
which sometimes suppress expected quantum phenomena
\cite{rempe, cirac, meekhof, brune}.
To discuss the effects of the noises, many attempts have been made 
to formulate the JC system interacting with the environment
especially with the GKSL-type master equation technique 
\cite{gks, lindblad}.

In this context, the JC system with cavity losses has been investigated 
theoretically in analytical and numerical ways over the years 
\cite{tranquang, barnett, puriagarwal, eiselt89, eiselt91, jgb},
most of which focused on collapses and revivals.
These publications typically treated the master equation of the form
\begin{equation}
	\dot{\rho}=-i[H_{\rm JC}, \rho] + 
	\gamma \left ( a\rho a^\dagger - \frac{1}{2} a^\dagger a \rho 
	- \frac{1}{2}\rho a^\dagger a \right),
\label{eq:masterph}
\end{equation}
where $\rho$ is the density matrix for the JC system that we focus on,
$H_{\rm JC}$  is the JC Hamiltonian,
$a^\dagger(a)$ is the field creation (annihilation) operator,
and $\gamma$ is the damping rate independent of the energy levels.
Since the JC system describes a Rydberg atom interacting 
with a resonant cavity of high quality factor and the photons are lost 
due to the imperfectness of the cavity, this master equation is apparently correct. 
However, this master equation is actually derived with an \textit{ad hoc} approximation,
and the microscopic process of the cavity losses is obscure.
The condition where this phenomenological master equation is justified is also unclear. 
Against this background, Scala \textit{et al.} \cite{scala,scala2} assumed a realistic microscopic 
interaction between the JC system
and the environment, derived a master equation for the resonant case 
using the standard technique \cite{breuer}, and analyzed the derived microscopic master equation. 
While the microscopic master equation gives a precise description, 
the expression is complicated and not easy to handle.
In particular, it is quite difficult to treat many-photon initial conditions analytically. 
Although the technique used in the derivation is also applicable to off-resonant case, 
the resulting master equation will be much more complicated.

In this paper, we present an alternative analytic method to discuss the JC system with 
cavity losses on the basis of the microscopic treatment.
We derive a simpler master equation by explicitly using the analytic solution to the 
JC model.
In this formalism, at zero temperature, we find the analytic solutions
for both the resonant and off-resonant cases under the approximation usually used.
The obtained solutions are suitable for the analyses not only for the 
single-excitation cases, which were examined by Scala \textit{et al.} \cite{scala}, 
but also for multi-excitation cases and the coherent state.
In particular, for the resonant case, we can write down the time evolution 
for general initial states explicitly.
Utilizing these solutions, we discuss some specific cases.
Furthermore, using the formula for finite detuning cases,
we examine the large and small detuning limits and discuss
the condition where the phenomenological treatment Eq.~\eqref{eq:masterph} is justified.

This paper is organized as follows.
In Sec.~\ref{sec:jc}, we review the properties of the pure JC model 
and rederive its analytic solution.
In Sec.~\ref{sec:mastereq}, using the obtained expressions, we derive a master equation
with the standard method.
In Sec.~\ref{sec:analsol}, we derive an analytic solution to the derived master equation 
for the resonant case, i.e., when the detuning $\Delta$ is zero and 
discuss the behavior of the initially multi-excitation states
including the coherent state.
In Sec.~\ref{sec:offres}, we derive an analytic solution for finite $\Delta$, 
discuss the large and small $\Delta$ limits,
and evaluate the time evolution 
for some specific cases.
Section~\ref{sec:summary} is devoted to discussions and summary.

\section{The Jaynes-Cummings model and its analytic solution
\label{sec:jc}}
\subsection{Energy eigenvalues and energy eigenstates}
We consider a two-level atom with energy separation $\hbar\omega_0$ 
and a quantized single mode of the electromagnetic field
with frequency $\omega$.
We denote the ground state and the excited state of the atom as 
$\ket{g}$ and $\ket{e}$, respectively.
We introduce an interaction between the atom and the quantized mode.
In the rotating wave approximation, the system is described by the JC Hamiltonian
\begin{align}
H_{\rm JC}= \frac{1}{2} \omega_0 \sigma_z + \omega (a^\dagger a + \frac{1}{2})
+\lambda(a \sigma_+ + a^\dagger \sigma_-), \quad(\hbar=1)
\end{align}
where $a^\dagger$ ($a$) creates (annihilates) a quantized mode of 
energy $\omega$ and $\sigma_+ = \ket{e}\bra{g}, 
\sigma_- = \ket{g}\bra{e}$, and $\sigma_z = \ket{e}\bra{e} - \ket{g}\bra{g}$
are the operators for the atom.
The magnitude of the interaction is represented by $\lambda$.
Using the relation $\sigma_z= 2\sigma_+ \sigma_- -1$,
we can rewrite the Hamiltonian as \cite{ackerhalt}
\begin{align}
H_{\rm JC}= \omega N + C,
\end{align}
where $N=a^\dagger a + \sigma_+ \sigma_-$ is the total excitation and 
$C=-\Delta \sigma_z + \lambda (\sigma_+ a + \sigma_- a^\dagger)$
is the remaining part of $H_{\rm JC}$.
The detuning $\Delta=\frac{1}{2}(\omega-\omega_0)$ represents the 
deviation of the photon energy from the energy separation of the atom.
Making use of the fact that the operators $N$ and $C$ commute, we obtain eigenstates and eigenvalues
\begin{align}
  &\ket{E_0}=\ket{g,0}, \nonumber \\
  &\qquad{\rm for} \, E_0=\Delta, \\
  &\ket{E_{n+}}=\cos \theta_n \ket{g,n} + \sin \theta_n \ket{e,n-1}, \nonumber \\
  &\qquad{\rm for} \, E_{n+}=n\omega + \varepsilon_n,  \\
  &\ket{E_{n-}}=\sin \theta_n \ket{g,n} - \cos \theta_n \ket{e,n-1}, \nonumber \\
  &\qquad{\rm for} \, E_{n-}=n\omega -\varepsilon_n,
\end{align}
with
\begin{align}
  &\varepsilon_n = \sqrt{\Delta^2 + n\lambda^2}, \\
  &\cos \theta_n= \frac{\lambda}{|\lambda|} \sqrt{\frac{\varepsilon_n+\Delta}{2\varepsilon_n}}, \quad 
  \sin \theta_n= \sqrt{\frac{\varepsilon_n-\Delta}{2\varepsilon_n}}.
	\label{eq:alpha-beta}
\end{align}
Note that $C$ is also diagonal with respect to the energy eigenstates, and
the corresponding eigenvalues are given by
\begin{align}
  &C \ket{E_0}=\Delta \ket{E_0}, \\
  &C \ket{E_{\pm n}}=\pm \varepsilon_n \ket{E_{\pm n}}.
\end{align}

\subsection{Time evolution in the Heisenberg picture}
We introduce operators in the Heisenberg picture
\begin{align}
  a(t)=e^{iH_{\rm JC} t} a e^{-iH_{\rm JC} t}=e^{-i\omega t}e^{iCt}a e^{-iCt}.
\end{align}
We can put time dependence only on the left or right
of the operator $a$, which makes its handling easy in the later procedure.
Rewriting the operator $e^{iCt}$ by Euler's formula, we obtain
\begin{align}
  a(t)=&e^{-i\omega t} \cos(\sqrt{C^2}t)a e^{-iCt} \nonumber \\
  &+e^{-i \omega t} \frac{C}{\sqrt{C^2}} \sin (\sqrt{C^2}t)a e^{-iCt},
	\label{eq:rep1}
\end{align}
where the sign of $C$ is expressed as $C/\sqrt{C^2}$.
Then, using the relation
\begin{align}
  f(C^2)a = a f(C^2-\lambda^2),\label{eq:ca-exchange}
\end{align}
we can put the time-dependent operators on the right of $a$ as
\begin{align}
  a(t)=&e^{-i\omega t} a \cos (\sqrt{C^2-\lambda^2}t) e^{-iCt} \nonumber \\
  &+ie^{-i\omega t} \frac{C}{\sqrt{C^2}} a \sin (\sqrt{C^2-\lambda^2}t) e^{-iCt}.
	\label{eq:rep2}
\end{align}
Finally, we use Euler's formula again, which yeilds
\begin{align}
  a(t)=P_+ a e^{i(-\omega + \sqrt{C^2 - \lambda^2}-C) t} 
  +P_- a e^{i(-\omega - \sqrt{C^2 - \lambda^2} - C) t},
	\label{eq:heisenberga0}
\end{align}
where
\begin{align}
  P_\pm = \frac{1}{2} \left(1 \pm \frac{C}{\sqrt{C^2}}\right)
\end{align}
are projection operators which project states depending on the positive 
or negative eigenvalues of $C$, i.e.,
\begin{align}
  &P_\pm \ket{E_{n\pm}}=\ket{E_{n\pm}},  \\
  &P_\mp \ket{E_{n\pm}}=0,
\end{align}
for $n\geq 1$ and,
\begin{align}
	&P_+ \ket{E_0} = \left\{ 
		\begin{array}{cl}
			\ket{E_0} & (\Delta \geq 0) \\
			0& (\Delta < 0) \\
		\end{array},
		\right.  \\
	&P_- \ket{E_0} = \left\{ 
		\begin{array}{cl}
			0 & (\Delta \geq 0) \\
			\ket{E_0}& (\Delta < 0) \\
		\end{array},
		\right.
\end{align}
for the ground state.
%
Note that other difinitions of $P_\pm$ on $\ket{E_0}$ are also possible when $\Delta=0$.
Similarly, we can apply the Euler's formula to $e^{-iCt}$ using the Hermitian conjugate of 
Eq.~(\ref{eq:ca-exchange}).
Then we obtain another expression, 
\begin{align}
  a(t)=&e^{i(-\omega+C-\sqrt{C^2+\lambda^2})t}aP_+ 
	  +e^{i(-\omega+C+\sqrt{C^2+\lambda^2})t}aP_-. \label{heisenberga}
\end{align}

\section{Microscopic derivation of the master equation
\label{sec:mastereq}}
In this section, we derive a master equation for the Jaynes-Cummings system
in the usual manner.
We assume the Hamiltonian for the environment and the interaction
between the Jaynes-Cummings system and the environment given by
\begin{align}
  &H_{\rm B}=\sum_k \omega_k b^\dagger_k b_k, \\
  &H_{\rm int}=(a+a^\dagger)\otimes B,
\end{align}
where $b_k^\dagger$ ($b_k$) is a creation (annihilation) operator for 
a boson in the environment with wave number $k$, 
$\omega_k$ is the energy of a boson, and 
$B$ is given by
\begin{align}
  B=\sum_k g_k (b_k+b^\dagger_k)
\end{align}
with $g_k$ characterizing the coupling between the JC system 
and the environment.
We denote the density matrices for the JC system, the environment,
and the total system as $\rho_{\rm JC}(t)$, $\rho_{\rm B}(t)$, and $\rho_{\rm tot}(t)$, 
respectively.
The first two density matrices are written in terms of the total density matrix as 
\begin{align}
  &\rho_{\rm JC}(t)=\tr_{\rm B} \rho_{\rm tot}(t), \\
  &\rho_{\rm B}(t)= \tr_{\rm JC} \rho_{\rm tot}(t),
\end{align}
where $\tr_{\rm B}$ and $\tr_{\rm JC}$ stand for partial traces taken over the degrees of freedom
for the environment and the Jaynes-Cummings system, respectively.

Here we assume that the environment is always in the thermal equiblium, i.e.,
\begin{align}
	\rho_{\rm B}(t)=\rho_{\rm B}\equiv e^{-\beta H_B}/\tr_{\rm B} e^{-\beta H_{\rm B}},
\end{align}
where $\beta=1/k_{\rm B}T$ with $k_{\rm B}$ and $T$ being 
the Boltzmann constant and the temperature, respectively.
The time development of the total system is described by the
von Neumann equation
\begin{align}
  \dot{\rho}_{\rm tot}(t)= -i[H_{\rm tot}, \rho_{\rm tot}(t)],
\end{align}
where $H_{\rm tot}=H_{\rm JC}+H_{\rm B}+H_{\rm int}$ is the 
total Hamiltonian.
After the Born and the Markov approximations,
we obtain a time evolution equation in the interaction picture
\begin{align}
  \dot{\rho}^{\rm I}_{\rm JC}(t)=-\int_0 ^\infty d\tau \tr_{\rm B}
  [H_{\rm int}^{\rm I}(t),[H^{\rm I}_{\rm int}(t-\tau),
  \rho^{\rm I}_{\rm JC}(t) \otimes \rho_{\rm B}]],
\end{align}
where the superscript $^{\rm I}$ denotes the interaction picture:
\begin{align}
  O^{\rm I}(t)=e^{i(H_{\rm JC}+H_{\rm B})t}Oe^{-i(H_{\rm JC}+H_{\rm B})t},
  \label{eq:interpic}
\end{align}
for an operator $O$ in the Schr\"{o}dinger picture.

Using the explicit form of $H_{\rm int}$
and tracing out the degrees of freedom for the environment, 
we obtain
\begin{align}
  &\dot{\rho}^{\rm I}_{\rm JC}(t) \nonumber \\
	&=-\int_0^\infty d\tau g(\tau)
  [a^{\rm I}(t)+a^{\rm I \dagger}(t)]
  [a^{\rm I}(t-\tau)+a^{\rm I \dagger}(t-\tau)]
  \rho^{\rm I}_{\rm JC}(t) \nonumber \\
  &\quad+\int _0^\infty d\tau g(\tau)
  [a^{\rm I}(t-\tau) + a^{\rm I \dagger}(t-\tau)]
  \rho^{\rm I}_{\rm JC}(t) 
  [a^{\rm I}(t) + a^{\rm I \dagger}(t)] \nonumber \\
  &\quad+ {\rm h.c.},
	\label{eq:masterinterim}
\end{align}
where $g(\tau) = \tr_{\rm B} [B^{\rm I}(\tau)B^{\rm I}(0)\rho_{\rm B}]$
is the two-time correlation function.
In each term of Eq.~(\ref{eq:masterinterim}), we can put together the time dependence of $a(t)$'s at one place
using Eqs.~(\ref{eq:heisenberga0}) and (\ref{heisenberga}) and the completeness condition,
\begin{equation}
	I=\ket{E_0}\bra{E_0} + \sum_{n=1}^\infty \sum_{\pm} \ket{E_{n\pm}}\bra{E_{n\pm}}
	\label{eq:completeness}
\end{equation}
on both sides of $\rho^{\rm I}_{\rm JC}(t)$ in the second term.
Carrying out the secular approximation, we obtain
\begin{align}
  \dot{\rho}^{\rm I}_{\rm JC}= 
  &-\frac{1}{2} \sum_\pm \{ P_\pm a^\dagger \gamma(\omega-C \pm \sqrt{C^2+\lambda^2}) a P_\pm,
  \rho^{\rm I}_{\rm JC}(t) \} \nonumber \\
  &-\frac{1}{2} \sum_\pm \{ P_\pm a \gamma(-\omega-C \pm \sqrt{C^2-\lambda^2}) a^\dagger P_\pm,
  \rho^{\rm I}_{\rm JC}(t) \} \nonumber \\
  &+\sum_{\pm}\gamma(\omega -C \pm\sqrt{C^2+\lambda^2}) \mathcal{P}^{\rm diag} 
  [a P_\pm \rho^{\rm I}_{\rm JC}(t) P_\pm a^\dagger] \nonumber \\
  &+\sum_{\pm}\gamma(-\omega -C \pm\sqrt{C^2-\lambda^2}) \mathcal{P}^{\rm diag} 
  [a^\dagger P_\pm \rho^{\rm I}_{\rm JC}(t) P_\pm a],
	\label{eq:master}
\end{align}
where
$\mathcal{P}^{\rm diag}$ is a projection operator for a density matrix onto the diagonal eigenbasis
with respect to the Hamiltonian,
which is defined by 
\begin{align}
  \mathcal{P}^{\rm diag} [\rho]=&
  \ket{E_0}\bra{E_0} \rho \ket{E_0}\bra{E_0} \nonumber \\
  &+ \sum_{n,\pm} \ket{E_{n\pm}}\bra{E_{n\pm}} \rho \ket{E_{n\pm}}\bra{E_{n\pm}}.
	\label{eq:diag}
\end{align}
The damping rate $\gamma(\Omega)$ is given by
\begin{align}
  \gamma(\Omega)
  =& \int_{-\infty}^\infty d\tau e^{i\Omega \tau}g(\tau) \nonumber \\
	=& 2\pi \sum_k g_k^2 [ \delta(\Omega-\omega_k) (N(\omega_k)+1) 
	+ \delta(\Omega+\omega_k) N(\omega_k)],\label{eq:gamma-omega}
\end{align}
where $N(\omega_k)=(e^{\beta \omega_k}-1)^{-1}$ is the average boson number of the environment,
and we have omitted the unitary part since it is usually negligibly small.
The argument $\Omega$ is an operator, 
and thus $\gamma(\Omega)$ is also an operator.
Note that the general formula Eq.~(\ref{eq:master}) is still valid for finite temperatures and 
finite $\Delta$.

\section{dynamics of the resonant JC system with cavity losses
\label{sec:analsol}}
\subsection{General formalism}
In the following, we consider the zero temperature cases and assume a cavity in the one-dimensional space
with an electromagnetic field.
This cavity is characterized by $g_k=\sqrt{c\gamma/2L}\,(\gamma>0)$,
where $c$ is the speed of light, $\gamma^{-1}$ is the lifetime of the cavity mode,
and $L$ is the size of one-dimensional space.
The operators $b$ and $b^\dagger$ are for photons, and the dispersion relation $\omega_k=c|k|$ is assumed.
Substituting this to Eq.~\eqref{eq:gamma-omega} and carrying out the $k$ summation, we obtain
\begin{equation}
  \gamma(\Omega)=
    \begin{cases}
      \gamma & (\Omega>0) \\
      0 & (\Omega<0)
    \end{cases}.
\end{equation}
This condition after all results in the effective Lorentzian coupling 
between the atom and the external electromagnetic field \cite{shimizu}.
We further assume $\omega \gg \lambda,\Delta$.
Then the excitation number-raising terms in Eq.~(\ref{eq:master}) 
(the second and fourth terms) vanish.
%
These conditions simplify the master equation Eq.~(\ref{eq:master}), and we obtain
\begin{align}
	\dot{\rho}^{\rm I}_{\rm JC}(t)&= 
	-\frac{\gamma}{2} \sum_\pm \{ P_\pm a^\dagger a P_\pm, \rho^{\rm I}_{\rm JC}(t) \} \nonumber \\
	&+\gamma \sum_\pm \mathcal{P}^{\rm diag}[aP_\pm \rho^{\rm I}_{\rm JC}(t) P_\pm a^\dagger]. \label{simplemaster}
\end{align}

Let us examine the dynamics described by Eq.~(\ref{simplemaster}) in the case of $\Delta=0$.
To solve Eq.~(\ref{simplemaster}), we decompose the density matrix $\rho^{\rm I}_{\rm JC}$ into 
the $\hjc$-diagonal and $\hjc$-off-diagonal sectors as follows:
\begin{align}
	\rho^{\rm I}_{\rm JC}(t)=\rho^{\rm I,diag}_{\rm JC}(t) + \rho^{\rm I,off-diag}_{\rm JC}(t),
\end{align}
with
\begin{align}
  \rho^{\rm I,diag}_{\rm JC}(t) &= \mathcal{P}^{\rm diag} [\rho^I_{\rm JC}(t)], \\
  \rho^{\rm I,off-diag}_{\rm JC}(t) &= \rho^I_{\rm JC}(t) -\mathcal{P}^{\rm diag} [\rho^I_{\rm JC}(t)].
\end{align}
Note that the equation is closed in each sector.
For each sector, 
the master equations are given by
\begin{align}
	&\dot{\rho}^{\rm I,diag}_{\rm JC}(t)=
  -\gamma A \rho^{\rm I,diag}_{\rm JC}(t)
	+ \gamma \mathcal{P}^{\rm diag} [a \rho^{\rm I,diag}_{\rm JC}(t) a^\dagger]. \label{eq:master-res-diag} \\
  &\dot{\rho}^{\rm I,off-diag}_{\rm JC} (t)
  = - \frac{\gamma}{2}(A\rho^{\rm I,off-diag}_{\rm JC}(t) 
  + \rho^{\rm I, off-diag}_{\rm JC}(t)A). \label{eq:master-res-offdiag}
\end{align}
with
\begin{align}
	A\equiv \sum_\pm P_\pm a^\dagger a P_\pm =N-\frac{1}{2}(1-P_0) \label{eq:large-a},
\end{align}
and $P_0=\ket{E_0} \bra{E_0}$. 
For the $\hjc$-off-diagonal sector, Eq.~(\ref{eq:master-res-offdiag}) is easily solved and we obtain the solution
\begin{align}
  \rho^{\rm I,off-diag}_{\rm JC}(t)
  =e^{-\frac{\gamma}{2}At}\rho^{\rm off-diag}_{\rm JC}(0)e^{-\frac{\gamma}{2}At}.
\label{eq:analsol-offdiag}
\end{align}

In the following, we concentrate on the $\hjc$-diagonal sector.
Applying the transformation \cite{honda}
\begin{align}
	\rho^{\rm I,diag}_{\rm JC}(t)
  =e^{-\gamma A t} \tilde{\rho}^{\rm I,diag}_{\rm JC}(t), \label{eq:tilde}
\end{align}
we obtain
\begin{align}
	\dot{\tilde{\rho}}^{\rm I,diag}_{\rm JC}(t)
  &=\gamma \mathcal{P}^{\rm diag}[e^{\gamma At}a e^{-\gamma At}
	\tilde{\rho}^{\rm I,diag}_{\rm JC}(t)a^\dagger] \nonumber \\
  &=\gamma \mathcal{K}_1(t) \tilde{\rho}^{\rm I,diag}_{\rm JC}(t), \label{deltazero1}
\end{align}
with $\mathcal{K}_1(t)$ being the linear operator for a density matrix defined by
\begin{align}
	\mathcal{K}_1(t)[\rho]=e^{-\gamma t(1-\frac{1}{2}P_0)}\mathcal{P}^{\rm diag}[a \rho a^\dagger], \label{deltazero2}
\end{align}
where we have used the relation $[a,A]=(1-{1\over2}P_0)a$.
The solution to Eq.~(\ref{deltazero2}) is formally written as 
\begin{align}
	\tilde{\rho}^{\rm I, diag}_{\rm JC} (t)
	= Te^{\gamma \int_0^t \mathcal{K}_1(t')dt'} \rho^{\rm diag}_{\rm JC}(0),
\end{align}
where $T$ represents the time-ordered product.
Since $aP_0=0$ and the product of $\mathcal{K}_1$'s at different times is written as 
\begin{align}
	\mathcal{K}_1(t_1) \mathcal{K}_1(t_2) \cdots \mathcal{K}_1(t_n)
	=e^{-\gamma (1-\frac{1}{2}P_0)t_1} e^{-\gamma(t_2+ \cdots + t_n)}(\mathcal{K}_2)^{n},
\end{align}
with
\begin{align}
	\mathcal{K}_2 [\rho]= \mathcal{P}^{\rm diag}[a\rho a^\dagger],
\end{align}
the ``commutation relation,''
\begin{align}
  &\mathcal{K}_1(t_1) \cdots \mathcal{K}_1(t_i) \cdots \mathcal{K}_1(t_j) \cdots \mathcal{K}_1(t_n) \nonumber \\
  &=\mathcal{K}_1(t_1) \cdots \mathcal{K}_1(t_j) \cdots \mathcal{K}_1(t_i) \cdots \mathcal{K}_1(t_n),
	\label{kequality}
\end{align}
holds for $1<i<j\leq n$.
Then, we can carry out the time-ordered product and we obtain
\begin{align}
	Te^{\gamma \int_0^{t} \mathcal{K}_1(t') dt'}
	=1+\gamma \int_0^t dt' e^{-\gamma (1-\frac{1}{2}P_0)t'}e^{(1-e^{-\gamma t'})\mathcal{K}_2}\mathcal{K}_2.
\label{eq:analsolint}
\end{align}
Performing the remaining integral, we obtain the time evolution of the $\hjc$-diagonal sector,
\begin{align}
	\rho^{\rm diag}_{\rm JC}(t)
	=e^{-\gamma At} 
	\left[ 1+ \sum_{n=0}^\infty c_n(t) (\mathcal{K}_2)^{n+1} \right]  
	\rho^{\rm diag}_{\rm JC}(0), \label{deltazerosol}
\end{align}
where
\begin{equation}
	c_n(t) = \sum_{k=0}^n \frac{(-1)^k}{k! (n-k)!} \frac{1-e^{-(1+k-\frac{1}{2}P_0)\gamma t}}{1+k-\frac{1}{2}P_0}.
\end{equation}

%
\subsection{Multi-excitation cases}
Our formulation Eq.~(\ref{eq:analsolint}) and Eq.~(\ref{deltazerosol}) are suitable for handling multi-excitation cases.
To consider the time evolution of these cases, it is convenient to see the behavior
of $\mathcal{K}_2$ acting on the $\hjc$-diagonal states.
Let us introduce a basis for the $\hjc$-diagonal sector of the density matrix,
\begin{align}
	&\Pi_{n\pm} = \ket{E_{n+}}\bra{E_{n+}} \pm \ket{E_{n-}}\bra{E_{n-}} \quad (n\geq 1) \nonumber \\
	&\Pi_0 = \ket{E_0}\bra{E_0}.
\end{align}
Then, we can show
\begin{align}
	&\mathcal{K}_2 \Pi_{n+}=\left( n-\frac{1}{2} \right) \Pi_{n-1,+} \quad (n\geq 2), \\
	&\mathcal{K}_2 \Pi_{n-}=\sqrt{n(n-1)} \Pi_{n-1,} \quad (n\geq 2), \\
	&\mathcal{K}_2 \Pi_{1+} = \Pi_0, \\
	&\mathcal{K}_2 \Pi_{1-} = \mathcal{K}_2 \Pi_0 = 0.
\end{align}
This recurrence equation is easily solved and we obtain
\begin{align}
	&(\mathcal{K}_2)^l \Pi_{n+}=\left(n-l+\frac{1}{2} \right)_l \Pi_{n-l,+}, \\
	&(\mathcal{K}_2)^l \Pi_{n-}=\sqrt{\frac{n!(n-1)!}{(n-l)!(n-l-1)!}}\Pi_{n-l,-},
\end{align}
for $n\geq l+1$, where $(n)_l$ is the Pochhammer symbol defined by
$(n)_l = n(n+1)\cdot \cdots \cdot (n+l-1)$.
Back in the original basis, we obtain
\begin{widetext}
\begin{align}
	(\mathcal{K}_2)^l \ket{E_{n\pm}}\bra{E_{n\pm}}
	=& \frac{1}{2} \left[ \left(n-l+\frac{1}{2}\right)_l 
	\pm \sqrt{\frac{n!(n-1)!}{(n-l)!(n-l-1)!}} \right] \ket{E_{n-l,+}}\bra{E_{n-l,+}} 
	\nonumber \\
	&+\frac{1}{2} \left[ \left(n-l+\frac{1}{2}\right)_l
	\mp \sqrt{\frac{n!(n-1)!}{(n-l)!(n-l-1)!}} \right] \ket{E_{n-l,-}}\bra{E_{n-l,-}} 
\end{align}
\end{widetext}
for $n\geq l+1$ and 
\begin{align}
	(\mathcal{K}_2)^n \ket{E_{n\pm}}\bra{E_{n\pm}}=\frac{(2n-1)!!}{2^n}
	\ket{E_0}\bra{E_0}
\end{align}
for $n=l(\geq 1)$.

Using these relations, we evaluate the time evolution $\rho_{gn}(t)$ for the multi-photon initial state
$\rho_{gn}(0)=\ket{g,n}\bra{g,n}$ with Eqs.\eqref{eq:interpic}, \eqref{eq:tilde}, and \eqref{eq:analsolint}, which is given by,
\begin{align}
	&\rho_{gn}(t) \nonumber \\
	&=\frac{1}{2} e^{-\gamma(n-\frac{1}{2})t}\Pi_{n+} \nonumber \\
	&\quad+\frac{1}{2} \sum_{m=1}^{n-1} e^{-\gamma(m-\frac{1}{2})t} 
	\frac{(1-e^{-\gamma t})^{n-m}}{(n-m)!} \left(m+\frac{1}{2}\right)_{n-m} 
	\Pi_{m+}\nonumber \\
	&\quad+\frac{(2n-1)!!}{(n-1)!2^n} B\left(n,\frac{1}{2};1-e^{\gamma t}\right)\ket{E_0}\bra{E_0} \nonumber \\
	&\quad+\frac{1}{2}e^{-\gamma(n-\frac{1}{2})t}(e^{-2i\sqrt{n}\lambda t}\ket{E_{n+}}\bra{E_{n-}}
	+\textrm{h.c.}) 
	\label{eq:manyphoton}
\end{align}
where
\begin{equation}
	B(a,b;z)=\int_0^z x^{a-1}(1-x)^{b-1} dx
\end{equation}
is the incomplete beta function.
Let us discuss the probability that we observe the ground state of the atom, $P_g$, and 
the average photon number $\langle n_{\rm photon}\rangle$, which are defined by
\begin{equation}
	P_g(t)=\sum_{k=0}^\infty \bra{g,k}\rho(t)\ket{g,k}, \label{eq:pgdef}
\end{equation}
and,
\begin{align}
	\langle n_{\rm photon} \rangle =& \sum_{k=1}^\infty k \bra{g,k}\rho(t) \ket{g,k} \nonumber \\
	&+\sum_{k=1}^\infty k \bra{e,k}\rho(t) \ket{e,k},\label{eq:nphotondef}
\end{align}
respectively. Substituting Eq.~(\ref{eq:manyphoton}) to Eqs.~(\ref{eq:pgdef}) and (\ref{eq:nphotondef}),
we obtain
\begin{align}
	P_g=&\frac{1}{2} + \frac{(2n-1)!!}{(n-1)!2^{n+1}}B\left(n,\frac{1}{2};1-e^{-\gamma t}\right) \nonumber \\
	&+ \frac{1}{2} e^{-\gamma(n-\frac{1}{2})t}\cos (2\sqrt{n}\lambda t),\label{eq:pg}
\end{align}
and
\begin{align}
	\langle n_{\rm photon} \rangle = &
	\frac{(2n-1)!!}{2^n(n-1)!}e^{-\gamma t/2}(1-e^{-\gamma t})^{n-1} \nonumber \\
	&\times {}_2F_1 \left(1,1-n,\frac{1}{2}; -\frac{1}{e^{\gamma t}-1} \right) \nonumber \\
	&+\frac{1}{2} e^{-\gamma(n-\frac{1}{2})t} \cos(2\sqrt{n}\lambda t),
\end{align}
where
\begin{equation}
	{}_2F_{1}(a,b,c;z)=\sum_{n=0}^\infty \frac{(a)_n (b)_n}{(c)_n}\frac{z^n}{n!}
\end{equation}
is the Gaussian hypergeometric function.
\begin{figure}[t]
\includegraphics[width=\linewidth]{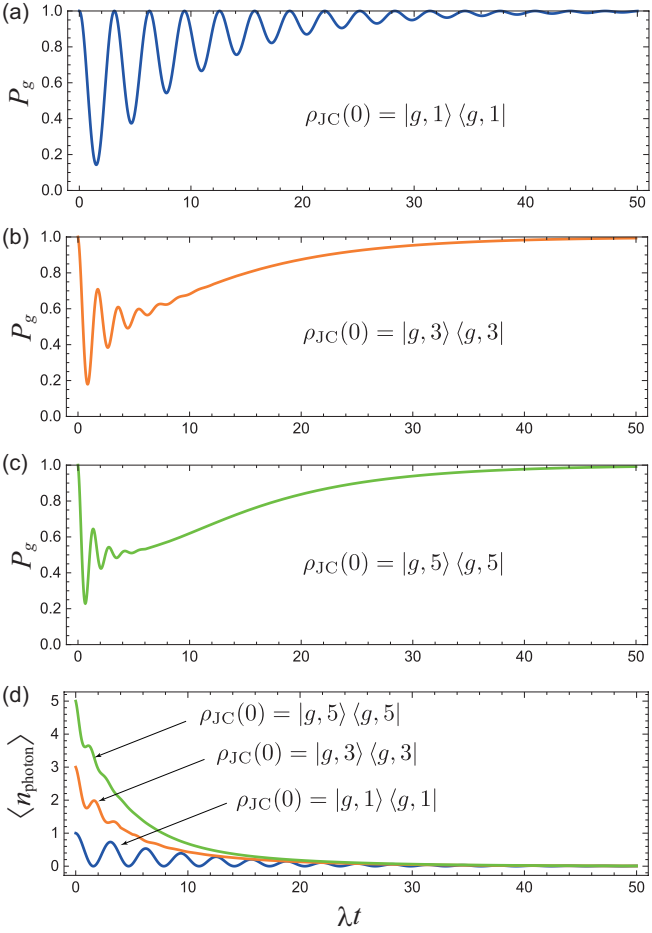}
\caption{(a)--(c): Probabilities of observing atomic ground state $P_g$ 
as functions of dimensionless time $\lambda t$ for the initial states
$\ket{n,g}\bra{n,g}$ with (a) $n=1$, (b) $n=3$, and (c) $n=5$.
(d): Averaged photon number $\langle n_{\rm photon} \rangle$ as functions of diemensionless time $\lambda t$
for the same initial states.
\label{fig:manyphotons}}
\end{figure}
Figure~\ref{fig:manyphotons}(a), (b), and (c) show $P_g$ for the initial states 
$\ket{1,g}\bra{1,g}$, $\ket{3,g}\bra{3,g}$, and $\ket{5,g}\bra{5,g}$, respectively.
There are several remarks on the above results.

(1) The period of the oscillation becomes shorter as the initial photon number increases.
$\ket{n,g}\bra{n,g} \,(n=1,3,5)$ contains the $\hjc$-off-diagonal part,
which oscillates with the period inversely proportional to the 
energy difference between the two energy eigenstates with the total excitation number $n$.

(2) The decay of the oscillation becomes faster as the initial photon number increases.
This is because the oscillatory part is suppresed by the factor $e^{-\gamma A t}$ typically,
and the eigenvalues of $A$ are almost proportional to the total excitation number [see Eq.~\eqref{eq:large-a}].

(3) The times that the initial states take to decay into the ground state $\ket{E_0}\bra{E_0}$ 
are not much different with the different initial photon numbers. 
However, as the initial photon number increases, the period for which $P_g$ stays around 0.5 becomes longer.

\subsection{Coherent state}
\begin{figure}[h]
\includegraphics[width=\linewidth]{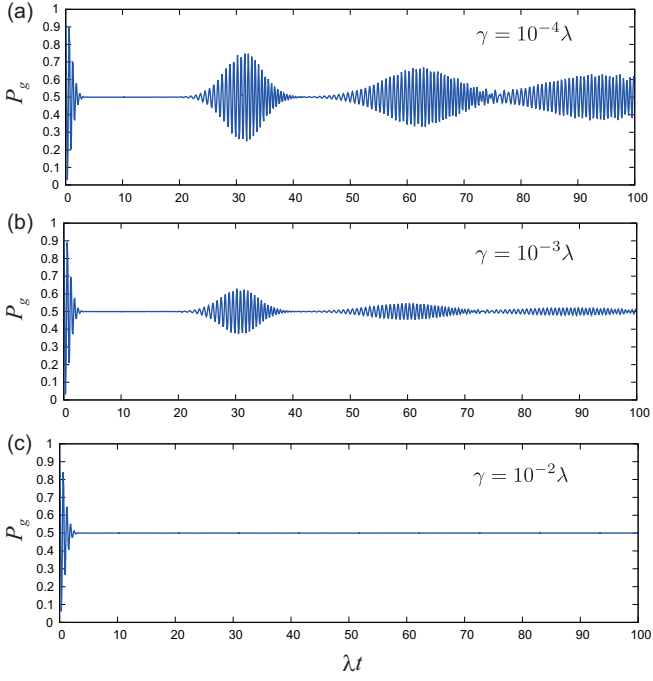}
\caption{Probability $P_g$ for the coherent initial state $\ket{g,\alpha}\bra{g,\alpha}$ 
is shown as a function of the diemensionless time $\lambda t$
for the case of (a) $\gamma/\lambda=10^{-4}$, (b) $\gamma/\lambda=10^{-3}$,
and (c) $\gamma/\lambda=10^{-2}$.
We set $\alpha=5$ (initial average photon number is 25).}
\label{fig:coherent}
\end{figure}

In this subsection, we consider the time evolution of the product state of photon coherent state
and atomic ground state, which is known to show collapses and revivals without cavity losses.
The photon coherent state is given by 
\begin{equation}
	\ket{\alpha}=e^{-\frac{\alpha^2}{2}}\sum_{n=0}^\infty \frac{\alpha^n}{\sqrt{n!}}\ket{n}
\end{equation}
where $\alpha$ is the parameter that characterizes the coherent state and we assume $\alpha$ is real 
here and in the following.
The average photon number is given by $\alpha^2$.
Using $\ket{\alpha}$, we set the initial state as
\begin{equation}
	\rho_\alpha(0)=\ket{g \alpha}\bra{g \alpha} 
	\equiv (\ket{g}\otimes\ket{\alpha})(\bra{g}\otimes\bra{\alpha}).
\end{equation}
We again decompose the density matrix into the $\hjc$-diagonal and $\hjc$-off-diagonal parts 
with respect to energy eigenstates.
We first consider the $\hjc$-diagonal part given by
\begin{align}
  \rho_\alpha^{\rm diag}(0)=e^{-\alpha^2}\ket{E_0}\bra{E_0} 
  + \frac{e^{-\alpha^2}}{2}\sum_{k=1}^\infty \frac{\alpha^{2k}}{k!} \Pi_{k+}.
\end{align}
Using Eqs.\eqref{eq:interpic}, \eqref{eq:tilde}, and \eqref{eq:analsolint}, after some algebra, we obtain
\begin{widetext}
\begin{align}
	\rho^{\rm diag}_\alpha(t)=&e^{-\alpha^2} \left[ 1+ \frac{1}{2}\sum_{n=1}^\infty 
	\frac{\alpha^{2n}}{(n-1)!n!} \frac{(2n-1)!!}{2^n} 
	B\left(n,\frac{1}{2};1-e^{-\gamma t}\right)\right] \Pi_0 \nonumber \\
	&+\frac{e^{-\alpha^2}}{2}\sum_{k=1}^\infty e^{-(k-\frac{1}{2})\gamma t}
	\left[ \frac{\alpha^{2k}}{k!} {}_1F_{1} \left( k+\frac{1}{2},k+1; \alpha^2(1-e^{-\gamma t})\right) \right] \Pi_{k+},
	\label{eq:coh-diag}
\end{align}
where 
\begin{align}
	{}_1F_1(a,b;z)=\sum_{n=0}^\infty \frac{(a)_n}{(b)_n}\frac{z^n}{n!}
\end{align}
is the confluent hypergeometric function.
The remaining $\hjc$-off-diagonal part is much easily treated and from Eqs.~\eqref{eq:interpic} and \eqref{eq:analsol-offdiag} we obtain
\begin{align}
	\rho^{\rm off-diag}_\alpha(t)=&\sum_{k=1}^\infty \frac{e^{-\alpha^2}\alpha^{2k}}{2\cdot k!} e^{-\frac{\gamma}{2}(2k-1)t}
	(e^{-2i|\lambda|\sqrt{k}t} \ket{E_{k+}}\bra{E_{k-}} + e^{2i|\lambda|\sqrt{k}t} \ket{E_{k-}}\bra{E_{k+}}),
	\label{eq:coh-offdiag}
\end{align}
where we have omitted the different--excitation number off-diagonal states since these terms do not affect $P_g$.
For this state, we obtain $P_g$ [Eq.~(\ref{eq:pg})] as follows:
\begin{align}
	P_g(t) = \frac{1}{2}+ \frac{e^{-\alpha^2}}{2}
	\left[ 1+ \frac{1}{2}\sum_{n=1}^\infty 
	\frac{\alpha^{2n}}{(n-1)!n!} \frac{(2n-1)!!}{2^n} 
	B\left(n,\frac{1}{2};1-e^{-\gamma t}\right)\right]
	+\sum_{n=1}^\infty \frac{e^{-\alpha^2} \alpha^{2n}}{2\cdot n!} e^{-\gamma(n-\frac{1}{2})t}
	\cos 2\lambda\sqrt{n}t,
\end{align}
\end{widetext}
which is shown in Fig.~\ref{fig:coherent} 
with $\gamma/\lambda=10^{-4}$, $10^{-3}$, and $10^{-2}$ cases.
With small $\gamma/\lambda$, we can observe clear collapses and revivals.
However, as the decay rate $\gamma$ becomes large, 
the long-time oscillation and the revivals are supressed more.
This result is consistent with a numerical calculation for the phenomenological master equation \cite{barnett}.

\section{dynamics of the off-resonant JC system with cavity losses
\label{sec:offres}}
In this section, we consider the off-resonant cases ($|\Delta|\neq0$)
and assume the same interaction with the cavity 
as Sec.~\ref{sec:analsol}.
After deriving an analytic solution,
we discuss the behavior of the time evolution for large and small $|\Delta|$ limits.
Also, we actually evaluate the time evolution for some initial conditions
using the obtained analytic solution
and compare the result with that derived from the phenomenological 
master equation.

\subsection{General formalism}
We start from Eq.~(\ref{simplemaster}).
The $\hjc$-off-diagonal dynamics is still described by Eq.~(\ref{eq:analsol-offdiag}) 
by substituting $\tilde{A}$ for $A$ with
\begin{equation}
  \tilde{A} =\sum_\pm P_\pm a^\dagger a P_\pm=N-\frac{1}{2} + \frac{\Delta}{2C}.
	\label{eq:a-tilde}
\end{equation}
For $\hjc$-diagonal states, on the other hand, we again perform a transform 
$\rho^{\rm I,diag}_{\rm JC}(t)=e^{-\gamma \tilde{A}t}\tilde{\rho}^{\rm I,diag}_{\rm JC}(t)$ 
and obtain the transformed master equation as follows:
\begin{align}
  \dot{\tilde{\rho}}^{\rm I \,diag}_{\rm JC} (t)
	=\gamma \mathcal{P}^{\rm diag}[e^{\gamma \tilde{A}t}a e^{-\gamma \tilde{A}t}
  \tilde{\rho}^{\rm I,diag}_{\rm JC}(t)a^\dagger].
\end{align}
Here we rewrite the time-dependent part as
\begin{align}
  &e^{\gamma \tilde{A}t} a e^{-\gamma \tilde{A} t}\nonumber\\
  &=e^{-\gamma t} e^{\gamma \frac{\Delta}{2C}t} a e^{-\gamma \frac{\Delta}{2C}t}\nonumber \\
  & =e^{-\gamma t} 
  [e^{-\frac{1}{2} \gamma \Delta t (\frac{1}{\varepsilon_{N+1}} - \frac{1}{C})}a P_+
  +e^{\frac{1}{2} \gamma \Delta t (\frac{1}{\varepsilon_{N+1}} + \frac{1}{C})}a P_-],
\end{align}
where we have used the relation obtained in a similar way to Eq.~(\ref{heisenberga}),
\begin{equation}
  a e^{-\gamma \frac{\Delta}{2C} t}= 
  e^{\frac{\gamma \Delta t}{2\varepsilon_{N+1}}}a P_- + e^{-\frac{\gamma \Delta t}{2\varepsilon_{N+1}}}a P_+.
\end{equation}
Then, the master equation for the $\hjc$-diagonal part reads
\begin{equation}
  \dot{\tilde{\rho}}^{\rm I,diag}_{\rm JC} (t)=
	\sum_{i=1,2}\gamma e^{-\gamma t \kappa_i(C)}\mathcal{Q}_i \tilde{\rho}_{\rm JC}^{\rm I,diag}(t), 
  \label{masterdelta}
\end{equation}
where
\begin{align}
  &\kappa_1(C)=1-\frac{\Delta}{2}\left(\frac{1}{C}-\frac{1}{\sqrt{C^2+\lambda^2}}\right), \\
  &\kappa_2(C)=1-\frac{\Delta}{2}\left(\frac{1}{C}+\frac{1}{\sqrt{C^2+\lambda^2}}\right), \\
  &\mathcal{Q}_1[\rho]=\mathcal{P}^{\rm diag}[aP_+ \rho P_+ a^\dagger],\\
  &\mathcal{Q}_2[\rho]=\mathcal{P}^{\rm diag}[aP_- \rho P_- a^\dagger],
\end{align}
The formal solution to Eq.~(\ref{masterdelta}) is given by,
\begin{align}
  \tilde{\rho}_{\rm JC}^{\rm I,diag}(t) = T \exp\left[\int_0^t \sum_i \gamma e^{-\gamma t' \kappa_i(C)}
  \mathcal{Q}_i dt'\right]\rho^{\rm I,diag}_{\rm JC}(0).
\end{align}
After the series expansion and the integration, we obtain
\begin{align}
  \tilde{\rho}^{\rm I,diag}_{\rm JC}(t)=\rho^{\rm I,diag}_{\rm JC}(0) 
  + \sum_{k=1}^\infty \mathcal{R}_k(t)[\rho^{\rm I,diag}_{\rm JC}(0)]
\end{align}
with
\begin{widetext}
\begin{align}
  \mathcal{R}_k(t)=\sum_{\{i_1, \dots,i_k\}=1,2} I_k(\gamma t; \kappa_{i_k}(\sigma_{i_{k-1}}\varepsilon_{N+k-1}),
  \dots, \kappa_{i_2}(\sigma_{i_1}\varepsilon_{N+1}), \kappa_{i_1}(C))
  \mathcal{Q}_{i_1}\cdots \mathcal{Q}_{i_k}
\end{align}
where each variable for the summation $i_1,\dots,i_k$ takes 1 or 2,
$\sigma_{i_j}$ is the sign: $+1$ for $i_j=1$ and $-1$ for $i_j=2$,
and $I_k(t;a_1,\dots,a_k)$ is the function determined by the recurrence formula,
\begin{align}
  &I_1(\tau;a_1)=\frac{1}{a_1}(1-e^{-a_1\tau}), \\
  &I_k(\tau;a_1,\dots,a_k)=\frac{1}{a_1}[I_{k-1}(\tau;a_2,a_3,a_4,\dots,a_n)-
  I_{k-1}(\tau;a_1+a_2, a_3,a_4, \dots,a_n)].
\end{align} 
The detail of the derivation is shown in the Appendix~\ref{sec:In}.
\end{widetext}

\subsection{Small and large $|\Delta|$ limits}
First, let us consider the case $|\Delta|\to 0$.
In this condition, Eq.~(\ref{eq:a-tilde}) reproduces Eq.~(\ref{eq:large-a}),
and thus the solution for $|\Delta|\to 0$ is smoothly connected to $\Delta=0$ case.

Next, we consider large $|\Delta|$ limit.
In the following, we assume $|\sqrt{n_\textrm{ave}}\lambda/\Delta| \ll1$,
where $n_\textrm{ave}$ is the average photon number of the initial state.
In this limit, $\cos \theta_{n}$ and $\sin \theta_{n}$ [defined in Eq.~(\ref{eq:alpha-beta})] approach unity and zero, respectively,
and $\ket{g,n}$ and $\ket{e,n-1}$ decouple.
Therefore, we can devide the master equation to the $\ket{E_+}(=\ket{g,n})$ part and 
the $\ket{E_-}(=-\ket{e,n-1})$ part:
\begin{align}
  \dot{\tilde{\rho}}^+_{\rm JC}(t)&= \gamma e^{-\gamma t \kappa_1(\varepsilon_N)}
	\mathcal{Q}_1[\tilde{\rho}^+_{\rm JC}(t)], \\
  \dot{\tilde{\rho}}^-_{\rm JC}(t)&=\gamma e^{-\gamma t \kappa_2(-\varepsilon_N)}
	\mathcal{Q}_2[\tilde{\rho}^-_{\rm JC}(t)], 
\end{align}
where $\tilde{\rho}^\pm_{\rm JC}(t)$ represents $P_\pm \tilde{\rho}^{\rm I,diag}_{\rm JC}(t) P_\pm$.
Furthermore, $\kappa_1(\varepsilon_N)$ and $\kappa_2(-\varepsilon_N)$ are evaluated as
\begin{align}
  \kappa_1(\varepsilon_N)=1+O\left(\left(\frac{\sqrt{N}\lambda}{\Delta}\right)^2\right), \\
  \kappa_2(-\varepsilon_N)=1+O\left(\left(\frac{\sqrt{N}\lambda}{\Delta}\right)^2\right). 
\end{align}
Hereafter, we assume that the total excitation number is small enough and we can neglect the 
higher order terms with respect to $|\sqrt{N}\lambda/\Delta|$,
which is regarded as the order of $|\sqrt{n_\textrm{ave}}\lambda/\Delta|$.
We expect that this condition is still maintained for the coherent state,
which contains infinitely many photons,
since there is an effective cutoff in the photon number determined by $\alpha$.
By this condition, we do not need to write $P_\pm$ in $\mathcal{Q}_1$ and $\mathcal{Q}_2$, 
and we obtain the master equations for $|\Delta|\to \infty$ limit as follows:
\begin{align}
  \dot{\tilde{\rho}}^\pm_{\rm JC}(t)=\gamma e^{-\gamma t} a\tilde{\rho}^\pm_{\rm JC}(t) a^\dagger.
	\label{eq:large-d-master} 
\end{align}
In this limit, we find $\tilde{A}\to a^\dagger a$, and 
the master equation in the Schr\"{o}dinger picture 
$\rho^{\pm}_{\rm JC}(t)(=e^{-\gamma\tilde{A}t}\tilde{\rho}_{\rm JC}^\pm)$ is expressed by
\begin{align}
	\dot{\rho}^\pm_{\rm JC}(t)=-{\gamma\over 2}a^\dagger a \rho^\pm_{\rm JC}(t)
	- {\gamma\over 2} \rho^\pm_{\rm JC}(t) a^\dagger a
	+\gamma a\rho^\pm_{\rm JC}(t)a^\dagger.
	\label{eq:large-d-master2}
\end{align}
In this form, we can readily confirm the trace-preserving property.
Similarly, we obtain the $\hjc$-off-diagonal part,
\begin{align}
	\dot{\rho}^{\rm off-diag}_{\rm JC}(t)=&
	-i[H_{\rm JC},\rho^{\rm off-diag}_{\rm JC}(t)] \nonumber \\
	&-{\gamma\over 2} a^\dagger a \rho^{\rm off-diag}_{\rm JC}(t)
	-{\gamma\over 2} \rho^{\rm off-diag}_{\rm JC}(t) a^\dagger a
	\label{eq:large-d-master-offdiag}
\end{align}
Master equations Eqs.~\eqref{eq:large-d-master} and \eqref{eq:large-d-master-offdiag} 
are immediately integrated, yielding
\begin{align}
  \rho^\pm_{\rm JC}(t)&=e^{-\gamma a^\dagger a t}e^{(1-e^{-\gamma t})\mathcal{K}_3}\rho^\pm_{\rm JC}(0),\nonumber \\
	\rho^{\rm off-diag}_{\rm JC}(t)&=e^{-iH_{\rm JC}t -{\gamma\over2}a^\dagger at}\rho^{\rm off-diag}_{\rm JC}(0)
	e^{iH_{\rm JC}t -{\gamma\over2}a^\dagger at},
\end{align}
with $\mathcal{K}_3\rho=a\rho a^\dagger$.

Through these analyses, we find two important facts.
First, Eqs.~\eqref{eq:large-d-master2} and \eqref{eq:large-d-master-offdiag} 
are the same as the phenomenological master equation Eq.~(\ref{eq:masterph})
apart from the jump term in the $\hjc$-off-diagonal part.
This fact suggests that the phenomenological master equation be justified for the 
case of $|\sqrt{n_{\textrm{ave}}}\lambda/\Delta|\ll1$.
This suggestion is confirmed for some examples in the following section.
Second, $\ket{E_{1-}}(=-\ket{e,0})$ does not decay to $\ket{E_0}$ 
because $\mathcal{K}_3\ket{E_{1-}}=0$.
This reflects the fact that $\ket{g,n}$ and $\ket{e,n-1}$ decouple 
in large $|\Delta|$ limit.
This effect is observed in the numerical evaluations to be shown later and Ref.~\cite{gonzalez}.

\subsection{Examples}
\subsubsection{Single excitation state}
\begin{figure*}[htb]
\includegraphics[width=\linewidth]{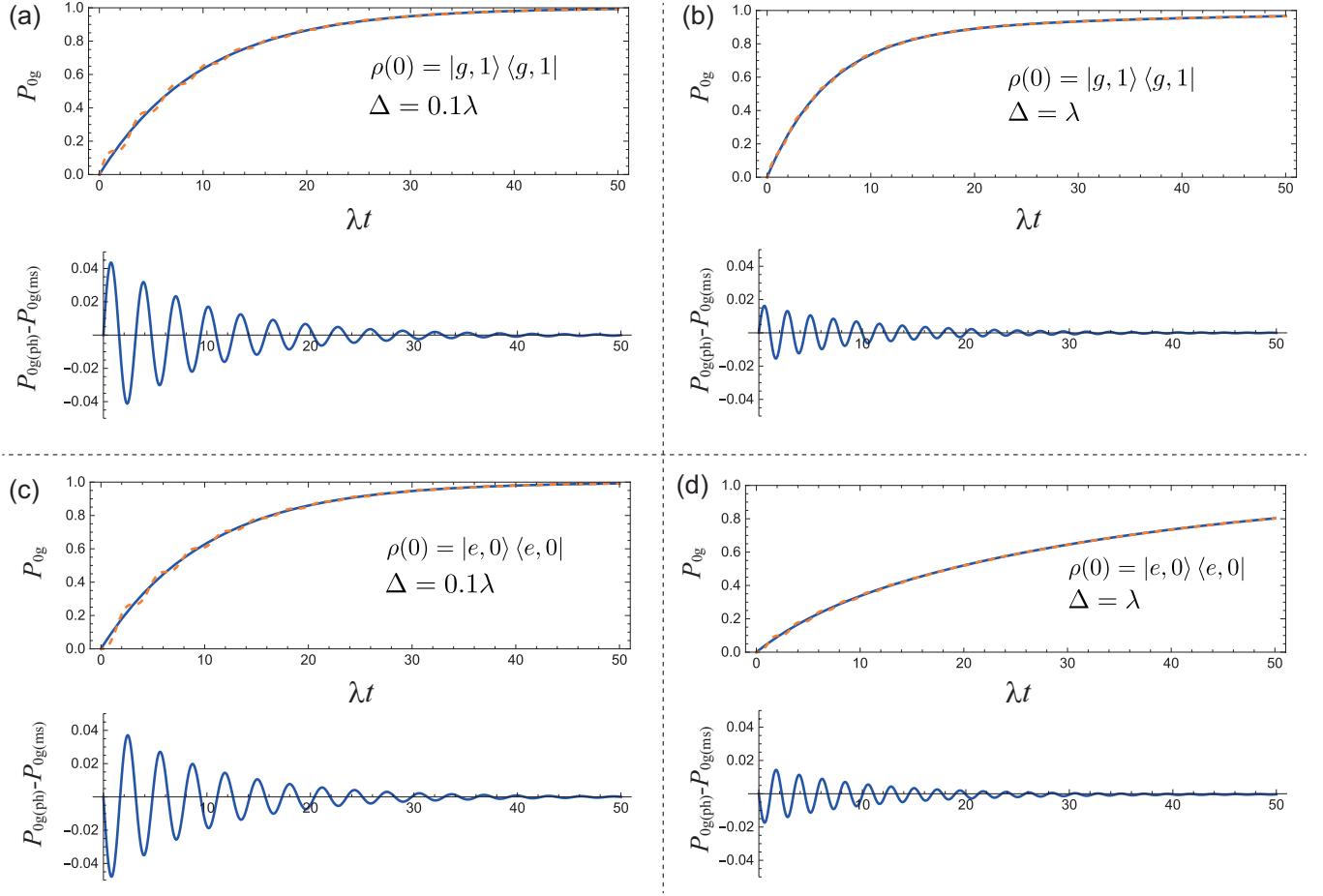}
\caption{Comparison between the probabilities for observing the ground state
evaluated by the microscopic (present) 
and phenomenological master equations.
The initial state and detuning are set to 
(a) $\ket{g,1}\bra{g,1}$ and $\Delta=0.1\lambda$,
(b) $\ket{g,1}\bra{g,1}$ and $\Delta=\lambda$,
(c) $\ket{e,0}\bra{e,0}$ and $\Delta=0.1\lambda$, and
(d) $\ket{e,0}\bra{e,0}$ and $\Delta=\lambda$.
The damping rate is set to $\gamma=0.2\lambda$ for all the cases.
The time evolutions are shown as functions of the diemensionless time $\lambda t$.
The solid (shown in blue, denoted by $P_{0g(\textrm{ms})}$) 
and dashed (shown in orange, denoted by $P_{0g(\textrm{ph})}$) lines represent the probabilities 
calculated from the microscopic and phenomenological master equations, respectively.
The differences for each set of time evolutions are also shown.}
\label{fig:jcdeltaex}
\end{figure*}
Using the obtained solution, we can immediately write down the time evolution for 
single-excitation initial states as follows.
\begin{widetext}
(i)$\rho_{\rm JC}(0)=\ket{E_{1+}}\bra{E_{1+}}$
\begin{align}
  \rho_{\rm JC}(t)&=e^{-\gamma \tilde{A}t}\left[\ket{E_{1+}}\bra{E_{1+}} 
  + \sum_{i=1,2} I_1(\gamma t;\kappa_i(C))\mathcal{Q}_i\ket{E_{1+}}\bra{E_{1+}}\right] \nonumber \\
  &=e^{-\gamma \tilde{A} t}\ket{E_{1+}}\bra{E_{1+}} + \frac{1}{2}\left(1+\frac{\Delta}{\sqrt{\Delta^2+\lambda^2}}\right)
  I_1(\gamma t;\kappa_1(\Delta))\ket{E_0}\bra{E_0} \nonumber \\
  &=e^{-\gamma \kappa_1(\Delta)t}\ket{E_{1+}}\bra{E_{1+}} + (1-e^{-\gamma \kappa_1(\Delta)t})\ket{E_0}\bra{E_0}
\end{align}

(ii)$\rho_{\rm JC}(0)=\ket{E_{1-}}\bra{E_{1-}}$
\begin{align}
  \rho_{\rm JC}(t)&=e^{-\gamma \tilde{A}t}\left[\ket{E_{1-}}\bra{E_{1-}} 
  + \sum_{i=1,2} I_1(\gamma t;\kappa_i(C))\mathcal{Q}_i\ket{E_{1-}}\bra{E_{1-}}\right] \nonumber \\
  &=e^{-\gamma \tilde{A} t}\ket{E_{1-}}\bra{E_{1-}} + \frac{1}{2}\left(1-\frac{\Delta}{\sqrt{\Delta^2+\lambda^2}}\right)
  I_1(\gamma t;\kappa_2(\Delta))\ket{E_0}\bra{E_0} \nonumber \\
  &=e^{-\gamma \kappa_2(\Delta)t}\ket{E_{1-}}\bra{E_{1-}} + (1-e^{-\gamma \kappa_2(\Delta)t})\ket{E_0}\bra{E_0}
\end{align}
\end{widetext}

Let us compare the time evolutions described by the microscopic master equation [Eq.~(\ref{masterdelta})] 
and phenomenological one [Eq.~(\ref{eq:masterph})].
We consider the Bell-type initial states $\ket{g,1}\bra{g,1}$ and $\ket{e,0}\bra{e,0}$ 
with finite $|\Delta/\lambda|$.
For these initial states, the $\hjc$-diagonal part is described by the linear combination 
of the results above, while the $\hjc$-off-diagonal part is described by Eq.~\eqref{eq:analsol-offdiag}
and \eqref{eq:a-tilde}.
The time evolutions of the probability 
that we observe the ground state $\ket{E_0}\bra{E_0}$, $P_{0g}$,
are shown in Fig.~\ref{fig:jcdeltaex}(a)--(d).
In the large $|\Delta|$ cases, we find two tendencies.
First, the time evolution desribed by the phenomenological master equation 
approaches to the that by the microscopic one.
These results supports the fact that the microscopic master equation almost coincedes with
the phenomenological one in the large $|\Delta|$ limit.
Second, we find that $\ket{e,0}\bra{e,0}$ decays much slower than $\ket{g,1}\bra{g,1}$ 
as mentioned by Gonzalez \textit{et al}. \cite{gonzalez}.
This is explained by the decoupling as follows.
While $\ket{g,1}\bra{g,1}$ can decay into $\ket{E_0}\bra{E_0}$ by operator $a$ directly,
$\ket{e,0}\bra{e,0}$ cannot decay, or rarely turns into $\ket{g,1}\bra{g,1}$ 
by the Rabi oscillation due to large $|\Delta|$.

\subsubsection{Three-photon state}
\begin{figure}[h]
\includegraphics[width=\linewidth]{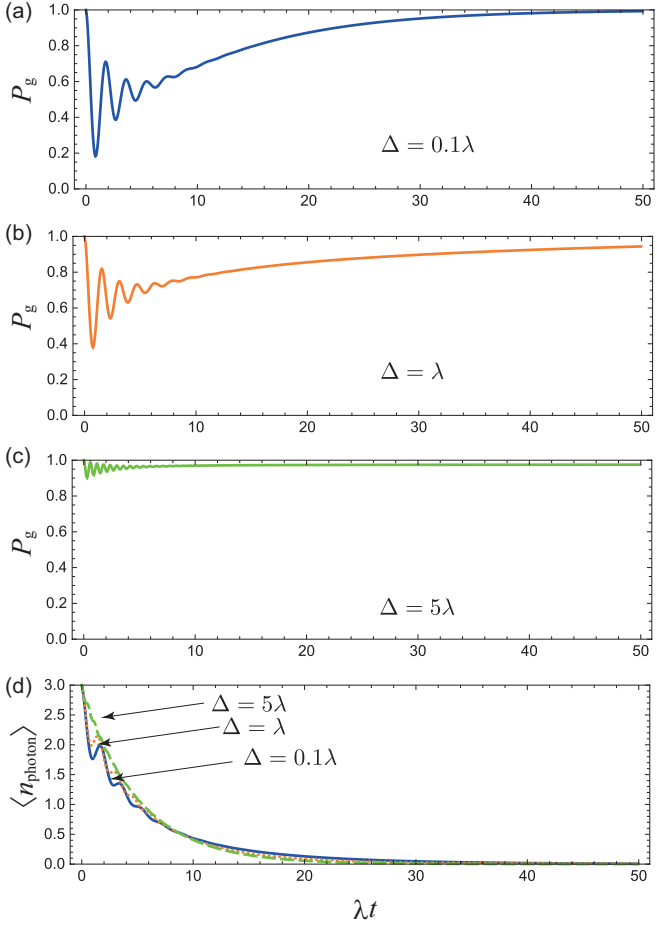}
\caption{(a)--(c): Probabilities $P_g$ for the initial state of three photons $\ket{g,3}\bra{g,3}$ 
as functions of the diemensionless time $\lambda t$
with (a) $\Delta=0.1\lambda$, (b) $\Delta=\lambda$, and (c) $\Delta=5\lambda$.
(d): Time evolution of the average photon number $\langle n_\textrm{photon} \rangle$.
Solid (blue), dot (orange), and dashed (green) lines correspond to the parameters for 
(a), (b), and (c), respectively.}
\label{fig:manyphoton-delta}
\end{figure}

Let us consider initial three-photon state as an example for multi-photon cases.
We fix the initial condition with $\rho_{\rm JC}(0)=\ket{g,3}\bra{g,3}$.
The time evolution of $P_g$ [Eq.~(\ref{eq:pg})] is shown in 
Fig.~\ref{fig:manyphoton-delta}(a) with $\Delta=0.1\lambda$,
Fig.~\ref{fig:manyphoton-delta}(b) with $\Delta=\lambda$, and
Fig.~\ref{fig:manyphoton-delta}(c) with $\Delta=5\lambda$.
As $|\Delta|$ becomes larger, the oscillation becomes faster.
This oscillation derives from the $\hjc$-off-diagonal part of $\rho_{\rm JC}(0)$,
and its period is dominated by the energy difference $2\sqrt{\Delta^2+3\lambda^2}$.
Furthermore, the oscillation is suppresed by the factor $e^{-\gamma \tilde{A}t}$ typically,
and the eigenvalues of $\tilde{A}$ is almost proportional to the total excitation number
[see Eq.\eqref{eq:a-tilde}].
Therefore, the oscillation is more suppresed with larger $|\Delta|$.

The average photon number is shown in Fig.~\ref{fig:manyphoton-delta}(d).
The solid (blue), dot (orange), and dashed (green) lines correspond to the condition 
for Figs.~\ref{fig:manyphoton-delta}(a), (b), and (c), respectively.
Apart from the fluctuation derived from the Rabi oscillation,
the average photon number decays to zero with the rate almost independent of $|\Delta|$.
\subsubsection{Coherent state}
\begin{figure}[ht]
\includegraphics[width=\linewidth]{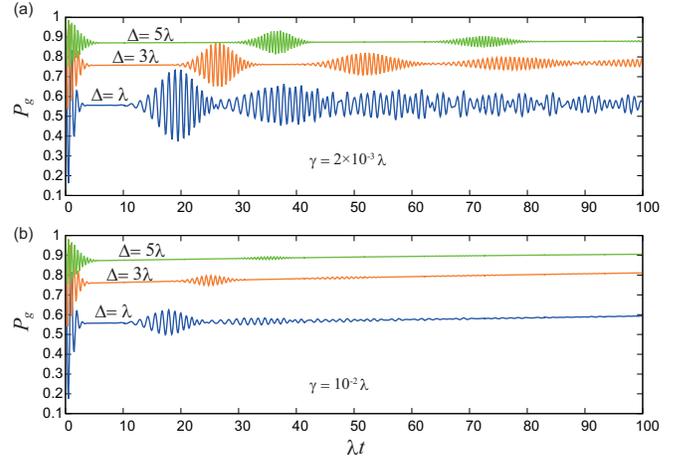}
\caption{Probability $P_g$ for the coherent initial state $\ket{\alpha,g}\bra{\alpha,g}$ 
is shown as a functions of the diemensionless time $\lambda t$
for $\gamma/\lambda=2\times10^{-3}$ and $10^{-2}$.
We set $\alpha=3$ (initial average photon number is 9).}
\label{fig:coherentdelta}
\end{figure}

Finally, let us consider the initial coherent state 
$\rho_{\rm JC}(0)=\ket{g,\alpha}\bra{g,\alpha}$.
Figure~\ref{fig:coherentdelta} shows the $P_g$ [Eq.~(\ref{eq:pg})] as functions of the diemensionless time $\lambda t$
for $\Delta=\{\lambda, 3\lambda,5\lambda\}$ and $\gamma=\{2\times 10^{-3},10^{-2}\}$.
We can see that the collapse-revival period becomes longer and the collapses and revivals 
become clearer as $|\Delta|$ becomes larger.
These tendencies are consistent with that observed in Gonzalez \textit{et al}.\cite{gonzalez}.
For larger $\gamma$, as in the case with $\Delta=0$, revivals are supressed faster
but $P_g$ takes relatively long time to decay into 1 compared with the case with a few photons 
presented in the previous subsection.

\section{discussions and summary
\label{sec:summary}}
Let us compare the present master equation with the previous result for the resonant ($\Delta=0$) case,
where the present result should be equivalent to Scala's one \cite{scala}.
This is shown as follows.
In the first two terms in Eq.~(\ref{eq:master}), we replace the projection operator $P_\pm$ by 
the sum of the projection operators for the eigenstates, $\ket{E_0}\bra{E_0}$, 
$\ket{E_{n+}}\bra{E_{n+}}$, and $\ket{E_{n-}}\bra{E_{n-}}$. 
Since the combinations $a^\dagger \gamma(\omega-C\pm\sqrt{C^2+\lambda^2})a$ 
and $a \gamma(-\omega-C\pm\sqrt{C^2-\lambda^2})a^\dagger$ do not change the total excitation number,
the excitation numbers of the resultant projection operators on the both sides should be the same.
Similarly, putting the completeness condition Eq.~(\ref{eq:completeness}) for both sides of $\rho_{\rm JC}^{\rm I}$
in the last two terms, we obtain an eigenstate-based expression.
After a lengthy but straightforward calculation, we obtain completely the same expression as Scala's.

It is also to be noted that we have assumed that the coupling between the cavity and the 
environment is flat through this paper.
This condition is not maintained for cavities in two- or three-dimensional space any more.
However, this condition is not the essential assumption for obtaining a closed form of solution.
When we set a specific coupling, 
the same discussion for the resonant or off-resonant cases in the present paper will be 
applied and we will be able to obtain expressions suitable for analytic analyses.

In summary, assuming a realistic coupling between the cavity and the environment,
we derived a master equation for the JC system with cavity losses.
The derived equation is simple, and we can write down the analytic solution explicitly
for the resonant case ($\Delta=0$) at zero temperature.
This solution is suitable for the analysis on many-photon states.
Using this solution, we clarified the many-photon effect on decay: 
The more photons exist, the slower decay rate of the atom becomes.
Also, we confirmed the clear collapses and revivals under dissipation.
For the off-resonant case, on the other hand, we developed an analytic way 
to describe the time evolution.
This method is systematic, although not explicit, and we can evaluate the time evolution similarly.
As examples, we examined the single- and multi-excitation and coherent initial states 
and revealed their various behaviors.
Also, we discuss the limits of $\Delta\to0$ and $\infty$,
and suggest the condition that justifies the widely-used master equation Eq.~\eqref{eq:masterph}:
$|\Delta/\lambda|$ is sufficiently large and the initial excitation number is small enough.
The present analytic methods are not only exact but also easy to handle 
in particular for the resonant cases
and will be useful for the analyses of experiments.

\begin{acknowledgments}
We thank A.\ Watanabe for fruitful discussions.
H.N. is partly supported by the Institute for Advanced Theoretical and
Experimental Physics, Waseda University and by Waseda University Grant
for Special Research Projects (Project No.2020C-272).
\end{acknowledgments}

\appendix*
\begin{widetext}
\section{Recurrence formula for $I_n$\label{sec:In}}
Let us first consider $\mathcal{R}_2(t)$.
The integral is given by
\begin{align}
	\mathcal{R}_2 (t) [\rho]
	=\sum_{\{i_1,i_2\}}\int_0^t dt_1 \int_0^{t_1} dt_2 \gamma^2 e^{-\gamma t_1 \kappa_{i_1}(C)} 
	\mathcal{Q}_{i_1}[e^{-\gamma t_2 \kappa_{i_2}(C)}\mathcal{Q}_{i_2}\rho].
\end{align}
Using the relation $af(N)=f(N+1)a$ for a function of the total excitation $f$,
we get the exponential terms out of $\mathcal{Q}$'s:
\begin{align}
	\mathcal{R}_2(t) [\rho]= \sum_{\{i_1,i_2\}}\int_0^t dt_1 \int_0^{t_1} dt_2 \gamma^2
	e^{-\gamma t_1 \kappa_{i_1}(C)} e^{-\gamma t_2 \kappa_{i_2}(\sigma(i_2)\varepsilon_{N+1})}
	\mathcal{Q}_{i_1}\mathcal{Q}_{i_2}\rho.
\end{align}
By the same procedure as above, we obtain
\begin{align}
	\mathcal{R}_k(t)[\rho]=\sum_{\{i_1, \dots,i_k\}} \int_0^t dt_1 \cdots \int_0^{t_{k-1}} dt_k
	\gamma^k e^{-\gamma t_1 \kappa_{i_1}(C)} e^{-\gamma t_2 \kappa_{i_2}(\sigma(i_i)\varepsilon_{N+1})}
	\cdots e^{-\gamma t_k \kappa_{i_k}(\sigma(i_k)\varepsilon_{N+k-1})}
	\mathcal{Q}_{i_1} \cdots \mathcal{Q}_{i_k}\rho.
\end{align}
Therefore, the integral we need to evaluate is of the form
\begin{align}
	I_n(t;a_1, \dots,a_n) \equiv  \int_0^t dt_n \int_0^{t_n}dt_{n-1} \cdots \int_0^{t_2}dt_1
	e^{-(a_n t_n+ \cdots + a_1 t_1)}.
\end{align}
For this integral, we find a recurrence relation,
\begin{align}
	&I_1(t;a_1)=\frac{1}{a_1}(1-e^{-a_1 t}), \\
	&I_n(t;a_1,\dots,a_n)=\int_0^t e^{-a_n t'} I_{n-1}(t';a_1,\dots,a_{n-1}) dt' \qquad(n\geq 2).
\end{align}
This relation is useful for numerical computation.
Furthermore, to obtain analytically handy expression, we explicitly calculate $I_2(t_2;a_1,a_2)$:
\begin{align}
	I_2(t;a_1,a_2)=\frac{1}{a_1} [I_1(t;a_2)-I_1(t;a_1 + a_2)].
\end{align}
Then we obtain, for example,
\begin{align}
	I_3(t;a_1,a_2,a_3)&=\frac{1}{a_1} \left[ \int_0^t e^{-a_3 t'}I_1(t';a_2)dt'
	-\int_0^t e^{-a_3 t'}I_1(t';a_1+a_2)dt'\right] \nonumber \\
	&=\frac{1}{a_1} [I_2(t; a_2,a_3)-I_2(t;a_1+a_2,a_3)],
\end{align}
or generally,
\begin{align}
	I_n(t;a_1,\dots,a_n)&=\frac{1}{a_1}[I_{n-1}(t;a_2,a_3,a_4,\dots,a_n)-I_{n-1}(t;a_1+a_2, a_3,a_4,\dots, a_n)].
\end{align}
\end{widetext}

\bibliography{jc}

\end{document}